\documentclass[aps,preprint,onecolumn,longbibliography]{revtex4-1}

\usepackage{graphicx,xcolor,booktabs}
\usepackage{gensymb}
\usepackage[lofdepth,lotdepth]{subfig}
\usepackage{ulem}

\usepackage{upgreek}
\newcommand{\micron}{$\mathrm{{\upmu}m}$}
\newcommand{\micronc}{$\mathrm{{\upmu}m^2}$}

\begin{document}


\title{Dynamics of Particle Migration in Confined Viscoelastic Poiseuille Flows}

\author{Antoine Naillon}
\email{antoine.naillon@univ-grenoble-alpes.fr}
\author{Cl\'ement de Loubens}
\author{William Ch\`evremont}
\author{Samuel Rouze}
\author{Marc Leonetti}
\author{Hugues Bodiguel}
\affiliation{Univ. Grenoble Alpes, CNRS, Grenoble INP, LRP, 38000 Grenoble, France}

\date{\today}

\begin{abstract}
Particles migrate in the transverse direction of the flow due to the existence of normal stress anisotropy in weakly viscoelastic liquids. We test the ability of theoretical predictions to predict the transverse velocity migration of particles in a confined Poiseuille flow according to the viscoelastic constitutive parameters of dilute polymers solutions.  Firstly, we carefully characterize the viscoelastic properties of two families of dilute polymer solutions at various concentrations using shear rheometry and capillary breakup experiments. Secondly, we develop a specific 3D particle tracking velocimetry method to measure with a high accuracy the dynamics of particles focusing in flow for Weissenberg numbers Wi ranging from $10^{-2}$ to $10^{-1}$ and particle confinement $\beta$ of 0.1 and 0.2. The results show unambiguously that the migration velocity scales as $\text{Wi}\beta^2$, as expected theoretically for weakly elastic flows of an Oldroyd-B liquid. We conclude that classic constitutive viscoelastic laws are relevant to predict particle migration in dilute polymer solutions whereas detailed analysis of our results reveals that theoretical models overestimate by a few tenth the efficiency of particle focusing.
\end{abstract}


\maketitle

\section{Introduction}\label{SecIntro}

When dispersed in a Newtonian liquid, isolated particles may exhibit cross-stream migration due to inertia \cite{HoL.G.1974} or to the deformability~\cite{Shevkoplyas2005}. In the semi-dilute regime, shear-induced migration also occurs~\cite{Acrivos1987}. However, no transverse forces are expected for dilute and non-deformable particles at low Reynolds numbers. In contrast, viscoelastic liquids exhibit normal stress anisotropy, which can lead to a net force on the particles in the transverse direction of the flow \cite{Ho1976}. One of the main interest of this force is that it is highly sensitive to the particle size since it scales as $a^3$, where $a$ is the particle size. This effect is promising to improve separation methods that are traditionally based on particle diffusivity \cite{Baalousha2011}. 

The consequences of viscoelastic lift force were first observed by Karnis and Mason~\cite{Karnis1966}, who reported that particles migrate in Poiseuille flow towards the centerline, where the shear is minimum. Ho and Leal \cite{Ho1976} and then Brunn \cite{Brunn1976} proposed a few years later an analytical prediction of this force, in the limit of inertialess flow, small confinement and weak fluid elasticity. Initially applied to plane Poiseuille flows, both group of authors extended their calculations to other geometries~\cite{chan1977, Brunn1980}. They obtained that the migration velocity is proportional to the Weissenberg number \text{Wi} and the square of the confinement ratio $\beta$. For a plane Poiseuille flow, they got a small difference of 10\% in the numerical prefactor~\cite{Brunn1980}. Thirty years later, Leshansky and coworkers~\cite{Leshansky2007} revisited this topic and evidenced viscoelastic particle migration in microchannels as a way to focus particles in the midplane of the flow. They reused a heuristic argument originally proposed by Karnis and Mason \cite{Karnis1966}, according to which the viscoelastic migration originates from the non-uniformity of the unperturbed flow at the scale of the particles. The gradient of normal stress on the particles leads to an unbalanced transverse force. Though in qualitative agreement with the prediction of Ho and Leal, one has to consider the perturbation of the flow due to the particle in order to achieve a quantitative prediction \cite{Ho1976}. 
Since the work of Leshansky and coworkers, the subject has gained particular interest and several groups of authors have been studying in more details the viscoelastic migration phenomenon in various flow geometries: square and rectangular microchannels \cite{Yang2011, Seo2014, DelGiudice2013}, pipes \cite{Romeo2013, Seo2014a}. The problem has also been extended to shear-thinning fluids \cite{Villone2011,DAvino2012} and deformable particles \cite{Yang2012,Go2017} by experiments or numerics. Finally, inertia-viscoelastic migration has also been investigated \cite{Lee2013,  Lim2014a}. As these various features fall out of the scope of the present article, we rather refer to four recent reviews \cite{DAvino2017, Yuan2018, Lu2017,Barbati2016}, which bring together a nice global overview of the subject. 

In summary, even in the simplest case of a inertialess plane Poiseuille flow, a quantitative comparison between theories and experiments has not been fully achieved yet, mainly due to two main experimental limitations detailed below. The main purpose of this work is to overcome these limitations and provide an accurate quantitative comparison. We believe that it will help the development of two types of applications that use the viscoelastic migration. In particle sorting ones (see e.g. reference \cite{Nam2015}), it is important to predict accurately the evolution of the probability density function (PDF) of particle position in the channel in order to better design devices. It has also been proposed \cite{DAvino2012} that one could use this effect to characterize the viscoelastic relaxation time of the solution. 

The first limitation for quantitative analysis is due to the difficulty to fully characterize the rheological properties. Most experiments are indeed carried out at low Weissenberg numbers, for which direct measurements of normal forces are not possible by standard rheometry methods. The fluid rheology characterization relies on the estimation of the viscoelastic relaxation time that is extrapolated from the loss and storage moduli. In addition, comparison with theory is based on an adequate viscoelastic model, which might not fully represent the system behavior. For instance, the characterization of fluid properties in \cite{Leshansky2007} leads to normal forces scaling as a power of 1-1.5 with respect to the shear rate whereas the various available constitutive equations \cite{Ho1976,Brunn1976} for viscoelastic fluids predict a power of 2. This prevents comparison with theory. In this paper, we directly measure the viscoelastic relaxation time of several solutions and check the conformity with classical constitutive viscoelastic laws. 

The second limitation is linked to the characterization of the phenomenon itself. At low Wi, migration velocities remain very small as compared to the transverse one, and it does not seem possible to measure it directly. To our knowledge, measurements are limited to consequences of the migration, and in particular, the PDF of particle position in the channel~\cite{Leshansky2007}. We reuse this strategy in this article, but with a greatly enhanced precision by measuring converged PDF along the channel, which usually requires at least a few thousands of particles. Romeo and coworkers \cite{Romeo2013, DelGiudice2013} used an inverse method based on velocimetry to determine particle positions in the cross section. However, the method requires to infer the flow profile, which is not straightforward for fluid with complex rheological properties or wall-slip effects. Also its uncertainty is rather large in the vicinity of the mid-plane, where the velocity gradient tends to zero. Seo and coworkers used a holographic method to determine the position of the particles in the channel \cite{Seo2014,Seo2014a} that does not require to scan the channel, in the direction normal to the focal plane. We adapt a similar microscopy technique for fluorescent particles. In addition to the above mentioned limitations, previous experimental works are limited to one or two fluids, and a systematic study on several polymer solutions is still lacking. 

We aim at quantitatively checking the analytical predictions of particle migration velocity in a viscoelastic fluid that were proposed forty years ago \cite{Ho1976,Brunn1976}, i.e. the migration velocity scales as $\text{Wi} \beta^2$. We specifically develop for that purpose a fast 3D particle tracking velocimetry to acquire PDF of the particle positions. The technique is applied at low Weissenberg numbers on two series of polymer solutions, which relaxation time is accurately determined using a capillary breakup experiment. This technique and the experimental details are described in the first section. The second section is devoted to the results and their discussion. In order to reach a quantitative comparison, the PDFs are systematically measured in different locations and their progressive thinning along the channel is predicted using the analytical expression of the transverse velocity.

\section{Materials and methods}\label{SecMaterialsMethods}
\subsection{Solutions of dilute polymers}\label{SSecMaterials}

Dilute polymer solutions are prepared from powders of polyacrylamide (HPAM, CAS 9003-05-8, $M_{n,HPAM}=150$ kDa) or polyvinylpyrrolidone (PVP, CAS 9003-39-8, $M_{n,PVP}=360$ kDa) that were provided by Sigma-Aldrich. They are dissolved in distilled water at a concentration of 4\% w/w with 1g/L of NaCl, filtered (0.2 \micron \ pore size syringe filters) and concentrated from 5\% to 11\% w/w by evaporation in an oven at 85 \degree C. After cooling, fluorescent polystyrene particles with a diameter of 9.9 \micron \ or 4.8 \micron \ (Fluoro-Max, Fisher Scientific) are dispersed in these solutions, with a volume fraction lower than 0.01\% so that interactions between particles are neglected. All the solutions properties are summarized in table \ref{TabManip}, together with experimental parameters, which will be detailed in the following.

\begin{table}[htbp]
 \centering
 \caption{Properties of the polymer solutions and experimental parameters. $n$ is the refractive index, values of the zero shear viscosity $\eta_0$ are in mPa.s, of $\tau$ in ms, and of $\dot{\gamma}_c$ and $\dot{\gamma}_{\text{max,exp}}$  in s$^{-1}$.}
 \begin{ruledtabular}
 \begin{tabular}{ccccccccc}
 polymer & \% w/w & $n$ & $\eta_0$ & $\tau$&  $\dot{\gamma}_c$ & $a$ & $\Delta P$ &  $\dot{\gamma}_{\text{max,exp}}$\\
 HPAM & 8 & 1.3467 & 148.9 & 1.1 & 258 & 4.95 & 100, 200, 300 & 105 \\
 HPAM & 9.5 & 1.35 & 368.8 & 2.7 & 211 & 4.95 & 100, 300, 500 & 81 \\
 HPAM & 10 & 1.3508 & 417.3 & 2.9 & 192 & 2.4 & 300, 500, 1000 & 127 \\
 HPAM & 11 & 1.352 & 657.2 & 3.6 & 140 & 4.95 & 100, 300, 500 & 64 \\
 PVP & 5 & 1.3414 & 40.0 & 1.6 & 369 & 4.95 & 75 & 130 \\
 PVP & 7.5 & 1.346 &133.2 & 3.5 & 132 & 4.95 & 100, 150, 200 & 78 \\
 PVP & 10 & 1.3502 & 311.5 & 6 & 100 & 4.95 & 100, 200, 400 & 67 \\
 \end{tabular}%
 \end{ruledtabular}
 \label{TabManip}%
\end{table}%

\subsection{Experimental set-up}

\begin{figure}[h]
\includegraphics[width=0.85\textwidth]{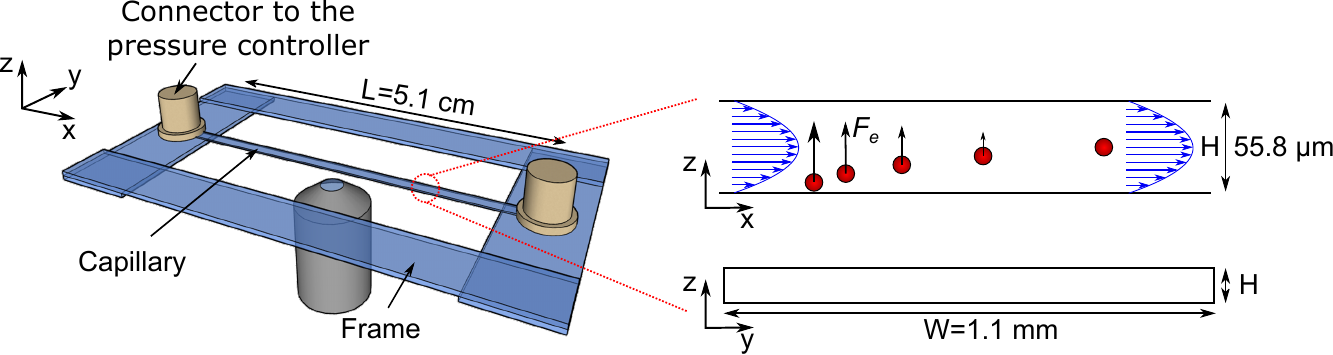}
\caption{Migration is observed in a thin rectangular micro-channel with an inverted microscope. Observation is made at $xy$ plane at different $x$ location.} \label{FigDispo}
\end{figure}

The experimental set-up consists in a thin rectangular glass capillary (CM Scientific) of $H \times w = 55.8 \times 1100 \ \mathrm{{\upmu}m^2}$ cross-section and $L=5.1 \ \mathrm{cm} $ length connected to the bottom of two connectors (Nanoport, Idex H\&S), Fig.~\ref{FigDispo}. The connectors are linked by PTFE tubes (inner diameter 1/16", Idex H\&S) to a syringe filled with a dilute polymer solution or to the atmospheric pressure. The flow is driven by a pressure controller (SMC ITV2010) connected to the syringe. The set-up is mounted on the 3-axis motorized moving stage (M\"arzh\"auser) of an inverted fluorescent microscope (Olympus ix73). The flow is observed in the small length direction by a 20x objective and images are recorded by a CMOS camera (Hamamatsu OrcaFlash 4.0). The field of view is $665\times665$ \micronc. 

Given the dimensions of the channel, the flow can be considered as bidimensionnal. The shear stress $\sigma$ can be directly calculated by $\sigma\left(z \right)=\frac{\Delta P}{L}z$ as we imposed the pressure drop $\Delta P$ between the inlet and outlet of the channel. $z=0$ is the mid-plane of the slit.

\subsection{3D-PTV: from 2D images to 3D particles trajectories}\label{SSecPTV}

As the flow is observed in the small length direction of the channel, particles migrate perpendicularly to the focal plane of the objective. In order to characterize this migration, a possibility would have consisted in scanning slice by slice the channel by moving the objective \cite{Leshansky2007}. Rather few particles per image would be detected in this case, leading to poor statistics or to a further increase of the acquisition time. An alternative is to infer the $z$-position from the velocity of the particles \cite{Romeo2013}. 

\begin{figure}[h]
\includegraphics[width=0.95\textwidth]{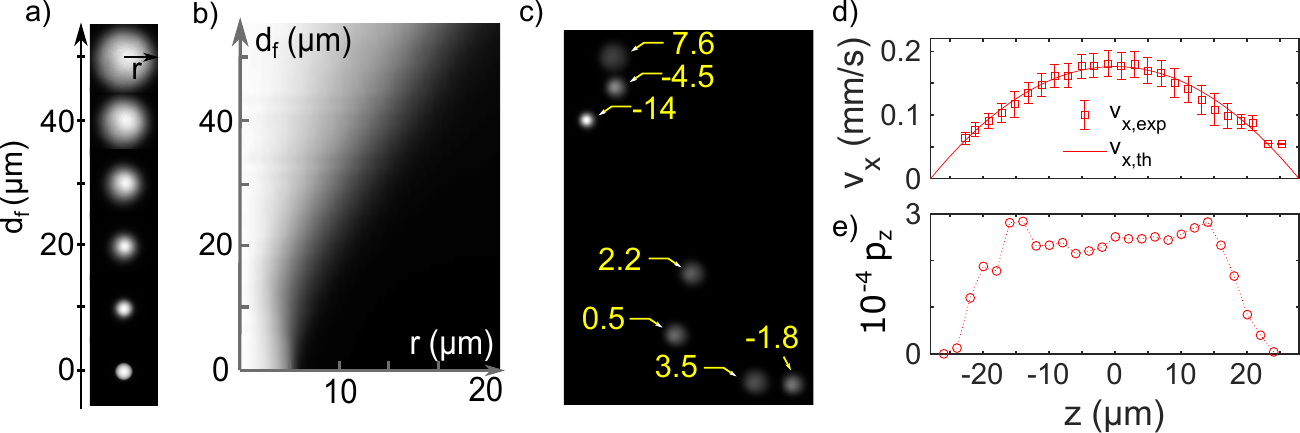}
\caption{ $z$-detection of particles. a) Fluorescence imaging of a particle at different distances $d_f$ to the focal plane. b) Calibration map obtained by orthoradial averaging of the images acquired at various $d_f$. c) Example of image obtained under flow. Numbers are the $z$ position ($z=d_f-H/2$) determined with the calibration map. d) Velocity profile determined by coupling PTV and $z$-detection. The solid line corresponds to the theoretical Poiseuille velocity profile. e) PDF of the $z$-position of the particles in Newtonian liquid at 1 cm of the inlet.} \label{FigPTVMethode}
\end{figure}

We have developed a method to measure 3D trajectories of particles in the whole cross section from 2D images acquired at a fixed position. It is based on a classical algorithm for particle tracking velocimetry (PTV) \cite{Crocker}, which is coupled with the determination of the $z$-position of particles. The latter takes advantage of the fact that the intensity profile of out-of-focus fluorescent particles directly depends on the distance to the focal plane \cite{Hiraoka1990}. As illustrated in Fig.~\ref{FigPTVMethode}a, the fluorescent signal emitted by one particle diffuses more and more when the distance between the particle and the focal plane of the microscope $d_f$ increases. Once calibrated, this effect allows us to determine the $z$-position of all the particles on each image. 

The calibration procedure consists in acquiring images of a fixed particle as a function of the distance to the focal plane, by moving the motorized objective of 1 \micron \ steps between each image. The recording lasts less than 10 seconds, and displacement due to sedimentation can be neglected on this time scale (the sedimentation velocity is about 1 nm/s, see section \ref{SSecResults}). Then, each image is normalized by its maximum as the emitted light could slightly differs from one particle to another\, and the radial intensity profile is computed. This procedure was done on several particles to get an averaged calibration map displayed in Fig.~\ref{FigPTVMethode}b. Note that the distance $d_f$ to the focal plane differs from the apparent distance $d_a$ of the objective displacement, as $d_f = d_a \times n_s/n_a $, $n_s$ and $n_a$ being the refractive indexes of the solution and air, respectively. They are measured by a refractometer Abbemat 350 (Anton-Paar) and reported in Table \ref{TabManip}. As the calibration map is symmetric with respect to the focal plane (not shown), the latter is placed on the bottom wall of the slit. With the 20x objective used in the study, we are able to locate particles which are up to 50 \micron \ out-of-focus, and thus to detect all particles that are flowing in the 50 {\micron}-height slit used. The image displayed in Fig.~\ref{FigPTVMethode}c shows an example of the application of the method. 

The $z$-detection procedure was validated by measuring the vertical position of fixed particles using the averaged calibration image (see supplemental materials \cite{supmat}, section I) and by extracting a Poiseuille flow profile for a Newtonian fluid. Coupling the 2D-PTV with the $z$-detection, we can effectively acquire the velocity profile in the slit cross section, Fig.~\ref{FigPTVMethode}d. The agreement with the theoretical profile is very good, without any fitting parameter.

\subsection{Probability density function measurment}\label{SSecPDF}

The probability density function $p_z$ of particle $z$-positions for a given location $x$ is defined as the probability to find a particle between the positions $z-\frac{\Delta z}{2}$ and $z+\frac{\Delta z}{2}$ divided by the class interval $\Delta z$ (2 \micron \ in this paper):
\begin{equation}
p_z(z) = \frac{1}{\Delta z}\frac{N(z-\frac{\Delta z}{2},z+\frac{\Delta z}{2})}{N_T}
\end{equation} 
where $N$ is the number of particles detected between $z-\frac{\Delta z}{2}$ and $z+\frac{\Delta z}{2}$, and $N_T$ the total number of detected particles. So that $\Sigma p_z \Delta z=1$

We measure $p_z$ by setting the microscope objective at a given $x$ position along the length of the channel. Sequences of 10 images are recorded at a high frame rate (typically 100 fps) in order to limit displacements of individual particles below a few pixels between consecutive images. This limits the number of particles that enter or exit the field of view during the acquisition. The average longitudinal velocity $v_x$ and $z$ position are calculated only for particles that are detected on at least 5 consecutive images. The delay between two image sequences is set long enough for all the particles to be out of the field of view in the following sequence. This procedure is repeated until we obtain around 3000 detected particles to get statistically independent PDF. For the experimental conditions used in this work, a full acquisition requires around 30 min.

The procedure has been first tested on the flow for a Newtonian fluid, and the result is displayed in Fig.~\ref{FigPTVMethode}e. One would expect uniformly distributed particles in a Newtonian liquid at low Reynolds numbers. The PDF indeed exhibits a large plateau in the central region of the channel, with small variations that are of the order of the measurement uncertainty. Near the wall, the PDF vanishes and we observe a depleted layer of approximately 8~$\mu$m. This layer is slightly larger than the particle radius. Thus, in addition to the excluded volume, there is another effect pushing particles away from the wall. As the depleted layer remains in this case much smaller than for particles dispersed in the polymer solutions, we did not put significant efforts to determine the physical origin of this small depletion, which could be due to the entrance region, or to interaction with the wall.

\subsection{Characterization of viscoelastic properties}\label{SSecRheoMethode}

The viscoelastic properties of the dilute polymer solutions are determined using two complementary methods, a rotational rheometer and an extensional one. 
Rotational rheometry is performed on a MCR 501 rheometer (Anton-Paar) with a cone and plate geometry (75 mm diameter, 0.991\degree \ angle). Steady-state viscosity $\eta$ is a priori dependent on the magnitude of the shear rates $\dot{\gamma}$. It is measured by increasing steps of shear rate from $\mathrm{10^{-1}}$ to $\mathrm{10^{3} \ s^{-1}}$ and decreasing steps of shear rate from $\mathrm{10^{3}}$ to $\mathrm{10^{-1} \ s^{-1}}$, Fig.~\ref{FigRheo}a (step duration is 30s). We did also small angle oscillations in the linear regime at various frequencies to measure the storage $G^{'}$ and loss moduli $G^{''}$ of the polymer solutions, Fig.~\ref{FigRheo}d.

To predict the transverse migration, the ideal characterization would consist in measuring also the $\mathrm{1^{st}}$ and $\mathrm{2^{nd}}$ normal stress differences, $N_1$ and $N_2$. However, for dilute polymer solutions, their values remain too low to be measured by standard rotational rheometers. The viscoelastic relaxation time is also difficult to determine using frequency sweep tests in the linear regime, as the upper frequency limit of the accessible range is not high enough to observe a crossover between $G'$ and $G''$. 

The viscoelastic relaxation times $\tau$ of the polymer solutions are measured with a capillary break up extensional rheometer (CaBER) using the slow retraction method~\cite{Campo-Deano2010} to avoid inertial effects. A droplet is trapped between two cylinders of diameter 4 mm, separated by a few millimeters. Then, one plate is moved very slowly, up to the destabilization of the liquid bridge due to capillary forces. As shown in Fig.~\ref{FigRheo}b, the liquid filament pinches, stretches and breaks up in a few tens of milliseconds, so that the gap could be considered fixed during this process. The experiments are recorded by a high speed camera Olympus i-speed 3 up to 5000 frames per second. To measure the relaxation time of the solution, we focus on the regime for which the two plates are connected by a cylindrical filament. The thinning of this filament is driven at late times by the competition between capillarity and elasticity. Assuming a cylindrical thinning of the filament in this regime, the radius $R$ decays exponentially with a characteristic time directly related to the viscoelastic relaxation time $\tau$ of the solution \cite{Campo-Deano2010,Entov1997,Clasen2006,AnnaShelley2001}: 
\begin{equation} \label{EqCapillary}
R \propto \exp\left({-t/3\tau}\right),
\end{equation}
As shown in Figure \ref{FigRheo}c, all the solutions tested exhibit such an exponential decay at late times. The values of $\tau$ deduced from the exponential fits are of the order of a few milliseconds.

\section{Results and discussion}\label{SecResults}

\subsection{Viscoelastic properties}\label{SSecRheoResultat}
	
\begin{figure}[]
\includegraphics[width=\textwidth]{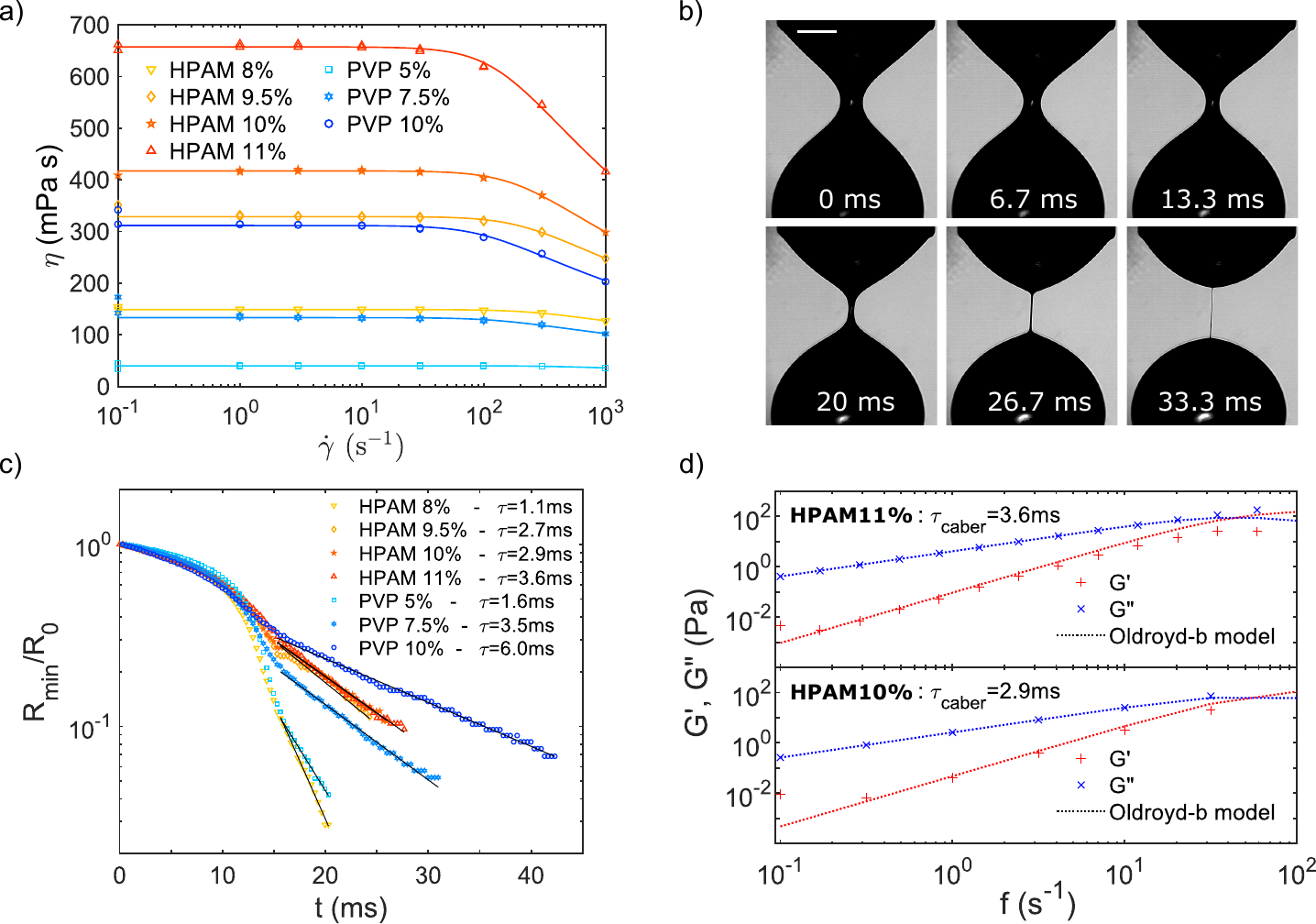}
\caption{Characterization of viscoelastic properties. a) Viscosity as a function of the shear rate (for each shear rate, one point is obtained for an increasing ramp, and one point for a decreasing ramp). Solid lines correspond to the best Carreau fits, used to determine the zero-shear viscosity of the solutions shown in table 1. b) Image sequence of capillary breakup (HPAM 10\%), white scale bar represents 1 mm. c) Evolution of the streched filament radius as a function of time, for all the solutions used. The solid lines correspond to the best exponential fits of the data. Eq~\ref{EqCapillary} is used to determine the relation time $\tau$. d) Linear oscillatory rheometry compared to the Oldroyd-B predictions (dotted lines) using the values of $\eta_0$ and $\tau$ determined by rotational rheometry and capillary breakup respectively. } \label{FigRheo}
\end{figure}

The shear viscosity $\eta$ of all the polymer solutions exhibits a Newtonian plateau at low shear rates and a shear thinning regime at high shear rates (Fig.~\ref{FigRheo}a). This behaviour is well approximated by a Carreau law, $\eta=\eta_0 \left(1+\left(\dot{\gamma}/{\dot{\gamma}_c}\right)^2\right)^{(n-1)/2}$ \cite{ByrondBird1968}, where $\eta_0$, $\dot{\gamma}_c$ and $n$ are the zero-shear viscosity, the critical shear rate that separates the two regimes, and the exponent of the shear-thinning, respectively. We assume in the following discussion that the polymer solutions tested have a constant viscosity since for the migration experiments, $\dot{\gamma}< \dot{\gamma}_c$ (see Table I). 

The zero-shear viscosity ranges from 40 mPa.s (PVP5\%) to 657 mPa.s (HPAM11\%), and the elastic relaxation time ranges from 1.1 ms (HPAM8\%) to 6 ms (PVP10\%), as determined using the capillary breakup experiments, see Fig.~\ref{FigRheo}. Interestingly, PVP solutions are less viscous than the HPAM ones for a given mass concentration whereas its elasticity is higher. This fosters a better efficiency for transverse migration (see Eqs.\ref{EqWi} and \ref{EqFeTh} later). 

Typical results obtained in linear oscillatory experiments are shown in Fig.~\ref{FigRheo}d. The loss and storage moduli exhibit a viscoelastic behavior, but the accessible frequency range lies in the viscous regime, as $G'' > G'$. These results are compared with the prediction of a Maxwell model, which is identical at small strains to the Oldroyd-B model \cite{Bird1987}, and reads
\begin{equation}
 G^{'}=\frac{\eta_0 \tau \omega^2}{1+\omega^2\tau^2} \ \ \text{and} \ \ G^{''}=\frac{\eta_0\omega}{1+\omega^2\tau^2}. \label{EqG}
 \end{equation} 
There is no fitting parameter in these expressions, as the values of the zero-shear viscosity $\eta_0$ and of the viscoelastic relaxation time $\tau$ are determined from the flow curves and the capillary breakup experiments, respectively. As shown in Fig.~\ref{FigRheo}d, the agreement is excellent, which proves the consistency of the measurements. It is noteworthy that though the viscoelastic relaxation times could have been estimated using extrapolations at high frequencies of the loss and storage moduli, the capillary breakup experiments lead to a much lower uncertainty and is much more adapted to relaxation times of the order of milliseconds. Furthermore, the storage moduli of the more dilute solutions fall below the accessible range of the rheometer.

As a conclusion, the polymer solutions are well modelled by the Oldroyd-B constitutive law and are compatible with the assumption of the theoretical development \cite{Ho1976,Brunn1976}, i.e. $N_1\sim\dot{\gamma}^2$.

\subsection{Dynamics of particle migration}\label{SSecResults}

\begin{figure}[h]
	\begin{center} 
		\subfloat[$c$=10\%, $a$=2.4\micron, $\Delta P$=0.3bar]{ 
		\includegraphics[width=0.45\textwidth]{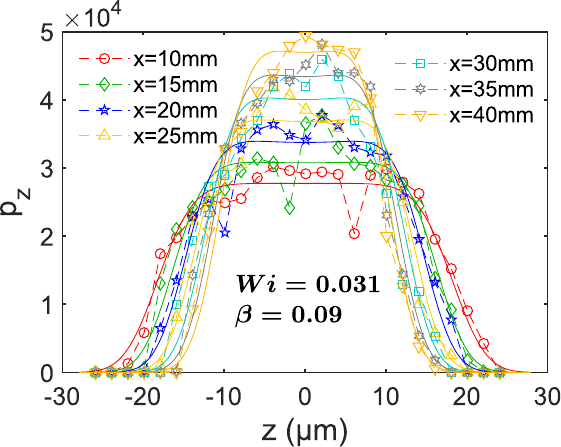} 
		\label{FigPDF1} } 
		\subfloat[$c$=10\%, $a$=2.4\micron, $\Delta P$=1bar]{ 
		\includegraphics[width=0.45\textwidth]{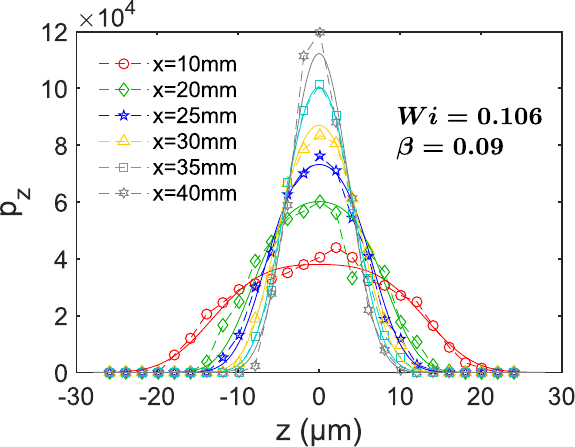} 
		\label{FigPDF2} }
		 
		\subfloat[$c$=9.5\%, $a$=4.95\micron, $\Delta P$=0.3bar]{ 
		\includegraphics[width=0.45\textwidth]{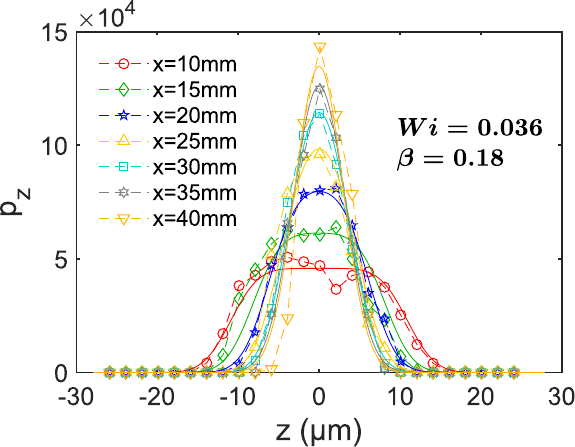} 
		\label{FigPDF3} } 
		\subfloat[$c$=11\%, $a$=4.95\micron, $\Delta P$=0.3bar]{ 
		\includegraphics[width=0.45\textwidth]{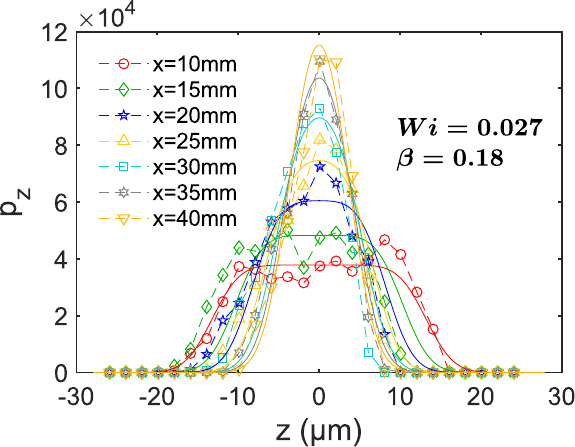} 
		\label{FigPDF4} } 
		\caption{Examples of evolution of the PDFs of the $z$-position along the flow direction $x$ (see legends). The higher the Weissenberg number $\text{Wi}$ and the confinement ratio $\beta$ are, the more efficient the focusing is. The solid lines are the best fits to the data (see text and appendix A).} \label{FigPDF} 
	\end{center} 
\end{figure}

Particle migration was determined by our 3D PTV algorithm. Typical PDFs of the $z$-position of the confined solid particles are displayed in Fig.~\ref{FigPDF} for different polymer concentrations, particle radii and pressure drops. For each case, the several PDFs shown are measured at different distances $x$ from the inlet of the channel. Clearly, the particles focus towards the mid-plane of the slit, as expected for viscoelastic migration. 

Let us first discuss whether the transverse particle migration can be attributed to other forces than elasticity. It could result from buoyancy forces, which lead to the Stokes velocity $v_s=\frac{2}{9}\frac{\Delta \rho}{\eta}a^2$, where $\Delta \rho$ is the difference of density between the liquid and the particle ($10\leq\Delta \rho \leq 35 \ \mathrm{kg/m^3}$ depending on the polymer solution). This gives $v_s\sim 1 \ \mathrm{nm/s}$. In the experiments, the longitudinal velocity is in the range $v_x\sim [0.1,1]$ mm/s, which means the residential time of a particle in the channel is $t_r=L/v_x\sim [500,50]$ s. Therefore, the displacement due to buoyancy along the channel is in the range [50-500] nm, and is always much smaller than the measured one. Buoyancy effects could therefore be neglected. Transverse migration could also be attributed to an inertial lift, the so called Segre-Silberberg effect \cite{Segre1961,HoL.G.1974}. Di Carlo reported that the migration velocity in a thin channel scales as $v_i \sim \frac{\rho v_x^2a^2}{H}$ or $v_i \sim \frac{\rho v_x^2a^5}{H^4}$ depending on whether the particle is near the mid-plane or near the wall \cite{DiCarlo2009}. Taking into account the maximum experimental velocity, this leads to $v_i\ < 1 $ nm/s and a migration displacement lower than 20~nm, again much smaller than the observed one. It can be concluded that elasticity is responsible to the transverse migration and other forces can be neglected.

Qualitatively, one can use Fig.~\ref{FigPDF1} and Fig.~\ref{FigPDF2} to see that the higher the pressure gradient, the faster the migration. Indeed, for a given system (HPAM solution, $c=10\%$, $a=2.4$ \micron), a better focusing is observed when increasing the pressure drop. The maximum of the PDF at the channel outlet is multiplied by a factor higher than 2 when the pressure drop is increased by a factor of 3. This trend is in agreement with the results of Leshansky et al. \cite{Leshansky2007}. It can also be observed that the focusing depends a lot on the particle radius, when comparing Fig.~\ref{FigPDF1} and \ref{FigPDF3}. For the same pressure drop ($\Delta P = 0.3$ bar), a slight change of rheological properties, and a particle radius multiplied by around two, the maximum of the PDF is increased drastically. At last, it is interesting to note the impact of a change in polymer concentration. Comparing Fig.~\ref{FigPDF3} and Fig.~\ref{FigPDF4}, increasing the polymer concentration reduces the focusing efficiency. This may appear counter-intuitive as the higher the polymer concentration, the higher the elasticity. In fact, as will be detailed below, the migration efficiency is governed by the Weissenberg number. The latter is defined as $\text{Wi} = \tau \dot{\gamma}_{\text{ref}}$, where $\dot{\gamma}_{\text{ref}}$ is a reference shear rate, which we define as $v_{\text{max}}/H$, where $v_{\text{max}}$ is the maximum velocity. Using Poiseuille law, this definition could be rewritten as 
\begin{equation} \label{EqWi}
\text{Wi} = \tau \dot{\gamma}_{\text{ref}}=\frac{H}{8}\frac{\tau}{\eta}\frac{\Delta P}{L}.
\end{equation}
For a given pressure drop, $\text{Wi}$ is proportional to the ratio $\tau/\eta$, which decreases when the concentration is increased in the range of the concentrations studied (see supplemental materials \cite{supmat}, section II). Thus, increasing the polymer concentration reduces the focusing efficiency for a given pressure drop.

\begin{figure}[]
\includegraphics[width=0.9\textwidth]{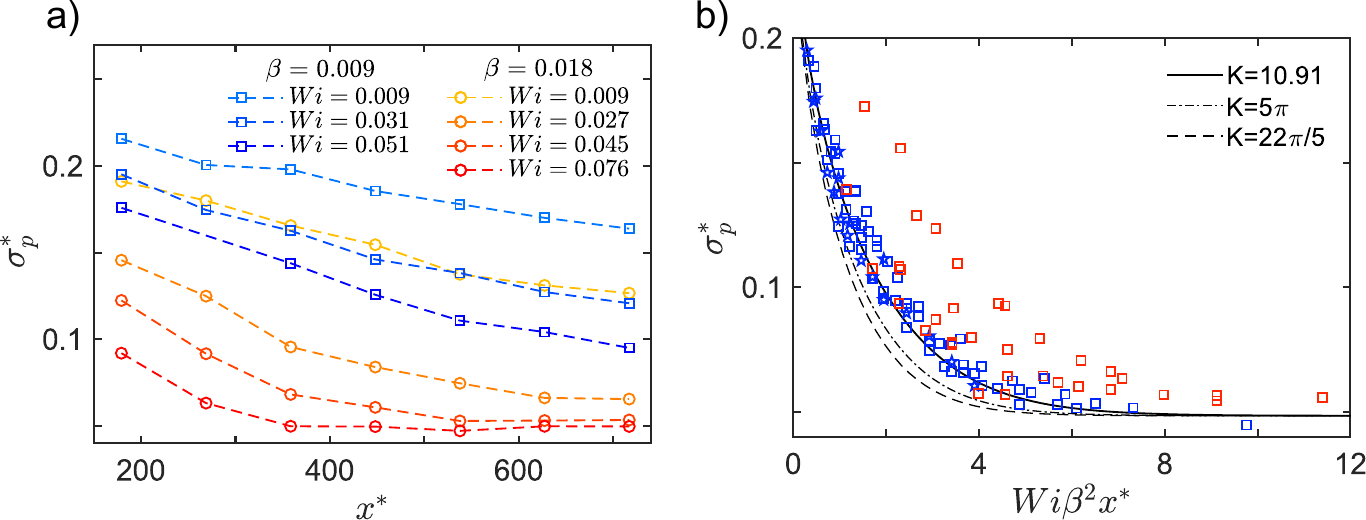}
\caption{ a) Standard deviation of the $z$-position of the particles $\sigma_p^*$, normalized by the channel height $H$, as a function of the normalized distance $x^* = x/H$ from the inlet, for various $\text{Wi}$ numbers. b) Master curve obtained by plotting $\sigma_p^*$ as a function of $\text{Wi}\beta^2 x^* $ taking the best fit of the prefactor K (see text) and the theoretical value coming from \cite{Ho1976} and \cite{Brunn1976}. All the data are gathered on the figure, red and blue symbols correspond to PVP and HPAM solutions, respectively.} \label{FigStd}
\end{figure}

Then, the migration efficiency is quantified by measuring the standard deviation $\sigma_p$ of the PDF of the $z$-position as a function of the normalized distance from the inlet $x^*=x/H$. As shown in Fig.~\ref{FigStd}a, the normalized standard deviation $\sigma_p^*=\sigma/H$ decreases for increasing $x^*$ and reaches a plateau of about 0.05 (2.7 \micron) in the most focused case. This value corresponds to the uncertainty of the $z$-position determination. 

According to \cite{Ho1976,Brunn1976}, the viscoelastic migration velocity scales as $\text{Wi} \beta^2$. This scaling is tested in Fig.~\ref{FigStd}b, where the normalized standard deviation of the particle $z$-positions is plotted as a function of $\text{Wi} \beta^2 x^* $. The collapse is excellent for the HPAM solutions and we note that data are more scattered for the PVP solutions. Let us underline that all the results are plotted Fig.~\ref{FigStd}b, and recall that not only the polymer and the concentration have been varied, but also the pressure drop and the particle size. Therefore, these results enable us to claim that the simple scaling in $\text{Wi} \beta^2$ is verified for small Weissemberg number and confinement lower than 0.2.

\subsection{Comparison with theory: PDF predictions and fitting} \label{SecComparison}

The aim of this section is to achieve test accurately theoretical laws: as explained in the introduction, an analytical approach based on perturbation theory was developed by Ho and Leal \cite{Ho1976} and Brunn \cite{Brunn1976} to model the particle migration velocity in a second order fluid. The corresponding constitutive law involves several fluid parameters that are not easy to determine independently, since the range of viscoelastic relaxation times of the solutions used in this work is below 10~ms. The second order fluid model leads to a constant viscosity in steady state shear flows and exhibits normal stress differences that scale as $\eta \tau \dot{\gamma}^2$. In fact, when the second normal stress difference is small, as it is usually the case in polymer solutions, the second order fluid model coincides with the Oldroyd-B one. In the following we assume that Oldroyd-B model could described well the viscoelastic properties of the polymer solutions, Figure \ref{FigRheo}. Under this assumption, the first normal stress difference is given by $N_1 = 2 \eta \tau \dot{\gamma}^2 $ and the second is zero. Rewriting the results of references \onlinecite{Ho1976} and \onlinecite{Brunn1976} with our notations and in the case of an Oldroyd-B fluid, we obtain that the transverse velocity is given by
\begin{equation} \label{EqFeTh}
v_z=\frac{K}{6\pi}\bar{v}_x \text{Wi}\beta^2 z^*,
\end{equation}
where $\beta=2a/H$ is the confinement ratio, $z^* =z/H$ the adimensional $z$-position and $\bar{v}_x$ is the mean longitudinal velocity. $K$ is a prefactor for which a slight discrepancy (10\%) exists in the literature. It equals $5\pi$ in Ho and Leal calculations~\cite{Ho1976}, and $22\pi/5$ in the paper of Brunn \cite{Brunn1976}. This expression justifies the existence of a master curve evidenced in Fig.~\ref{FigStd}b, where the standard deviation of the particle position is plotted as a function of $\text{Wi}\beta^2 x^* $. The fact that this scaling is well verified in the experiments indicates that the normal stress difference is proportional to $\eta\tau\dot{\gamma}^2$, in agreement with the Oldroyd-B model.

In order to achieve a quantitative comparison with the above prediction, we calculate the PDFs of the particle position as a function of the longitudinal position, and compare to the experimental ones. Calculation details are explained in Appendix. Note that we take into account the uncertainty of the measurement, by convolving the calculated PDFs with a Gaussian function. When trying to compare quantitatively the experimental data with the predictions, we find that the focusing was in most of the situations theoretically faster than observed. We thus decide to adjust the numerical prefactor $K$ appearing in the expression of the transverse force (Eq.~\ref{EqFeTh}), but keeping the scaling unchanged. For each experimental condition, we fit simultaneously the PDFs at different $x^*$ positions. The agreement is in all cases very good, as could be seen in the examples displayed in Fig.~\ref{FigPDF}. The best fits to the data are obtained with values of $K$ of about 10, ranging from 5 to 16, depending on the experiment. More precisely, the mean value is 10.9, and the standard deviation is 2.76. 

The values of $K$ are plotted in Fig.~\ref{FigK}, as a function of $\tau/\eta$, which only depends on the tested solutions. Data align quite well around a mean value, but the scatter does not exhibit any particular trend. By varying the pressure drop for a given solution, $K$ does not change significantly. As already noticed in Fig.~\ref{FigStd}, the data for the PVP solutions are more scattered than for the HPAM solutions. 	

Using the mean value of $K = 10.9$, we use the theoretical PDF to predict (after convolution) the standard deviation as a function of $x^*$ and compare to the experimental data displayed in Fig.~\ref{FigStd}. The agreement with the experimental master curve is excellent for the HPAM solutions, and reasonable for the PVP solutions. Note that in this case, and contrary to the fitting procedure, the PDF was assumed to be a uniform distribution at the inlet. This global agreement evidences the consistency of the experimental approach and of the analysis.

\subsection{Comparison with theory: discussion}

Strikingly, the numerical prefactor is 30\% lower than the theoretical prediction of Ho and Leal (5$\pi$) \cite{Ho1976}, and 20\% lower than that of Brunn ($22\pi/5$) \cite{Brunn1976, Brunn1980}, and this discrepancy is significantly greater than the scatter in the data. Let us now discuss this difference. 

We assume in this work that the rheological properties of the polymer solutions tested are accounted by the Oldroyd-B model as supported by our rheological data, which is a particular case of the second order fluid model with $N_2 = 0$ and $N_1 = 2 \text{Wi }\eta \dot{\gamma} $. The theoretical results of references \onlinecite{Ho1976} and \onlinecite{Brunn1976} are more general than the one considered in this work, and the transverse viscoelastic force defined in Eq.~\ref{EqFeTh} is multiplied by the factor $\left(1-2N_2/N_1\right)$ when $N_2$ is non zero. As $N_2$ is generally negative for polymer solutions, though much smaller than $N_1$, the theoretical prefactor $K$ is in fact slightly greater than 5$\pi$ (or than 22$\pi$/5). Thus, a potential contribution of $N_2$ cannot explain the difference with the theoretical prediction. 

Slippage at the wall can in contrast reduce the efficiency of the viscoelastic migration. Indeed, the velocity in the flow direction would be higher than expected for a given shear stress. Though the transverse velocity would remain constant, its effect on migration would appear weaker. However, the velocity profiles that are aquired concurrently with the PDF do not exhibit any slippage at the wall, and remains within experimental error well described by the expected Poiseuille velocity profile. 

It might also be argued that the steady-state viscosity is not perfectly constant, as a shear-thinning behaviour is observed at high shear-rate. This question is of great importance as previous studies reported that the viscoelastic migration is significantly modified in the case of shear-thinning fluids: particle migrates either towards the wall either towards the center, depending on their initial position \cite{DAvino2012}. However, we have explored a range of shear rates for the migration experiments that remains below the shear-thinning regime. Even for the most concentrated solutions (HPAM 11\% and PVP 10\%) for which the shear-thinning regime starts for $\text{Wi} \sim 0.2 $, which is close to the experimental range [$10^{-2}$,$10^{-1}$], we did not see any deviation from Poiseuille law, looking at the velocity profiles. For the less concentrated ones, the Newtonian plateau holds until $\text{Wi} \sim 1 $, and the migration experiments are conducted in flows that are at least one decade lower than this value. Since the discrepancy between the experiments and the theory is similar for both concentrated and dilute solutions, we think that shear-thinning is probably not the reason for it. 

Direct measurements of the normal force would be interesting to check that the Oldroyd-B model accounts well for the steady state large shear viscoelastic properties of the solutions. However, it is not possible using standard rheometry in the small range of $\text{Wi}$ tested. Nevertheless the fact that the experimental data collapse when plotted as a function of $\text{Wi}\beta^2 x^*$ indirectly indicates that the transverse velocity is proportional to the Weissenberg number, and the normal forces to $\dot{\gamma}^2$, as accounted by the Oldroyd-B model (and others). Moreover, the shapes of the PDFs change significantly if one considers normal forces proportional to $\dot{\gamma}^n$, with $n\neq 2$. As shown in the supplemental materials section III \cite{supmat}, the PDFs exhibit an important peak at the midplane in case $n>2$ and symmetric double peaks in case $n<1$. We were not able to fit well the experimental PDFs using a value of $n$ that is different from 2 ($\pm 0.2$).

\begin{figure}[h]
\includegraphics[width=0.5\textwidth]{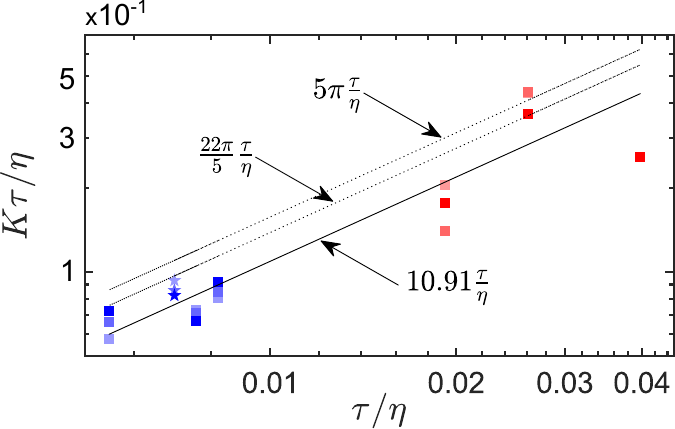}
\caption{Values of the effective compliance $K\tau/\eta$, where the prefactor $K$ is the best fit parameter obtained by fitting all the PDFs for a given experiment (same solution, particle and pressure drop). Blue symbols correspond to the HPAM solutions and red symbols to the PVP ones. Color darkness is related to the pressure drop, the darker the higher. Stars for the small particles ($\beta\simeq 0.1$), and squares for the biggest ($\beta\simeq 0.2$). The two dotted lines correspond to the two available theoretical predictions \cite{Ho1976,Brunn1976} while the solid line is the best linear fit to the experimental data, of slope 10.91. } \label{FigK}
\end{figure}

It is also interesting to recall the assumptions of Ho and Leal calculations \cite{Ho1976}. It is only valid in the limit of low $\text{Wi}$, which is the case here, but of unconfined particles, as it is assumed that $\beta \ll 1$. This last assumption might not be fully satisfied as we have tested $\beta \simeq 0.2 $ and $\beta \simeq 0.1 $ conditions. D'avino and coworkers have tested the role of confinement in a pipe \cite{DAvino2012} and reported that deviation from the $\beta^2$ regime occurs from $\beta > 0.11$. Above this value, the viscoelastic migration is weaker than the extrapolated one. A 10\% reduction is typically obtained for the highest reported value, $\beta \simeq 0.15$. It is thus temptative to conclude that the confinement might explain the 30\% difference that we observed, as most of the experiments have been conducted at $\beta = 0.2$. However, since the present experimental data are compatible with the $\beta^2$ scaling and that the $\beta= 0.1$ condition (stars in Fig.~\ref{FigK}) remains at least 20\% below the theoretical migration, we think that confinement is not likely to explain quantitatively the discrepancy.

Let us finally mention that other experimental details might also be considered, such as very small variations in the slit aperture, and particle asphericity, roughness or polydispersity. Although all of these are very small in the experiments, their consequences can be non-negligible since transverse migration is itself a small phenomenon. A small polydispersity could for example lead to a greater standard deviation of the $z$-position and would thus reduce the apparent migration. All of these imperfections would however be at least in a first approximation be equivalent to an uncertainty in the $z$-position of the particles. Yet, the experimental PDFs are accounted by a convolution with a Gaussian of standard deviation equal to the measurement uncertainty itself, excluding an additional contribution of the small experimental details mentioned above. 

To conclude this discussion, we were not able to identify clearly the reason why the migration appears about 20\% or 30\% weaker than theoretically predicted for unconfined particles in a second-order (or Oldroyd-B) fluid. Up to our knowledge, previous experimental work on viscoelastic migration did not report a similar deviation. However, as detailed in the introduction, these were probably not accurate enough to detect this rather small discrepancy. Extending the experiments to a lower range of $\text{Wi}$ and to less confined particles would be interesting to clarify the points discussed above. This would require a dedicated and more complex experimental setup to quantify the migration as it would ask for much longer slits.



\section{Conclusion}

We have developed a fast method of 3D-PTV, which does not require scanning on the third dimension. It allowed us to quantify with unprecedented precision and statistics viscoelastic particle migration in a confined Poiseuille flow. Not less than 7 different polymer solutions have been used at different Weissenberg numbers. The confinement ratio has also been tested. Importantly, the viscosity and the viscoelastic relaxation time have been determined precisely and independently to be able to test the available theoretical predictions. The results reported in this article unambiguously show that the viscoelastic transverse velocity is proportional to $\text{Wi}\beta^2 z^*$, as expected for small confinement ratio and Weissenberg numbers. Besides, the approach proposed allows to predict the evolution of the PDFs of the particle position along the channel, which we believe is interesting in particle sorting applications where viscoelastic migration could be used. 

Despite a good agreement with the scaling relation, the experimental results differ by a factor of 20\% or 30\% from the theoretical predictions, which overestimates the migration. We were not able to provide a satisfying explanation for this small discrepancy, which can originate e.g. from deviations to the rheological model or from the confinement ratio. We believe that for some applications, for example to use this phenomenon to characterize fluid viscoelastic properties, it is important to determine the prefactor that should be used, and this asks for the understanding of the difference between theory and experiments. We think that additional experimental efforts are required, dedicated on the one hand on extending the range of Weissenberg numbers and confinement explored in this work, and on the other hand on the fine characterization of the polymer solutions used. We think the methodology we have developed could be of great interest in this context as it allows a robust and precise characterization of the migration phenomenon. Numerical simulations might also bring some insight on the problem, notably one may be able to verify the analytics, to include confinement effects, and to incorporate more complex rheological properties and geometries. Perspectives of this work also deal with enhancing the accuracy of the particle position determination, as the PDF shapes can be used to infer the rheological of the fluids, provided that the measurement uncertainty is reduced below the particle size. Since we notice that the efficiency of particle focusing is increased when reducing the concentration, it would also be of interest to study more dilute solutions, in order to look for an optimum concentration.

\appendix*

\section{Calculation of theoretical PDFs and fitting procedure }
\label{appendix}

Rewriting the results of references \onlinecite{Ho1976} and \onlinecite{Brunn1976} with our notations and in the case of an Oldroyd-B fluid, we obtain that the transverse viscoelastic force acting on the particle is given by 
\begin{equation} \label{EqFeTh2}
F_{e,th}=- K \eta\bar{v}_xa \text{Wi}\beta^2 z^*,
\end{equation}
Next, balancing the viscoelastic transverse force with the drag force $6\pi\eta a v_z$ leads to 
\begin{equation} \label{EqVz1}
v_z=\frac{K}{6\pi}\bar{v}_x \text{Wi}\beta^2z^*
\end{equation}
where $z^*$ corresponds to the dimensionless $z$-position normalized by the length scale $H$. For a Poiseuille flow, the longitudinal velocity can be expressed as:
\begin{equation}
v_x=6\bar{v}_x\left( \frac{1}{4}-z^{*2}\right)
\end{equation}
and the particles trajectories can be deduced writing that $dz/dx=v_z/v_x$. It leads to
\begin{equation} \label{EqTrajTh}
x^*-x_0^*=\frac{18\pi}{K}\frac{1}{\beta^2 \text{Wi}}\left[\frac{1}{2}\ln\left(\frac{z^*}{z_0^*}\right)-\left({z^*}^2-{z_0^*}^2\right)\right],
\end{equation}
where ($x_0$,$z_0$) is the initial position of the particle.

Knowing the trajectories of the particles, it is possible to predict the PDFs along the channel axis. For that purpose, we define $p$ as the global PDF of the $z$-position of the particles. $p$ differs from the measured PDF $p_z$, which is defined for a given position $x$ in the channel, whereas $p$ refers to the probability of finding a particle at a position $(x,z)$. One can write $p (x,z) = \phi(x) p_z(x,z) $, where the unknown function $\phi(x)$ can easily be deduced by stating that $\int p_z dz =1$, i.e. $\phi(x)= \int p dz $. The interest of the global PDF $p$ comes from the fact that it respects the continuity equation, given in steady state by $\mathbf{\nabla}\cdot \left(\mathbf{v} p\right) =0,$ where $\mathbf{v}$ is the velocity vector of the particles. A simplification of this equation arises because both the longitudinal component $v_x$ of the velocity and the transverse one $v_z$ depend solely on $z$. It could be rewritten as:
\begin{equation} \label{EqConservation}
 v_x \partial_x q + v_z \partial_z q = 0, 
 \end{equation}
where $q=v_z p$. This last equation is a homogeneous quasilinear PDE, which characteristic lines coincide with the particle trajectories. Therefore, $q=v_z p$ is constant along the trajectories. This result provides a simple way to compute the PDF $p$, knowing $v_z$ (Eq.~\ref{EqVz1}) and the particle trajectories (Eq.~\ref{EqTrajTh}), and assuming a boundary condition $p(x_0,z_0)$ at $x=x_0$. It is given by 
\begin{equation} \label{EqPz}
 p(x,z) = \frac{v_z\left(z_0 \right)}{v_z(z)} p\left(x_0,z_0\right), 
\end{equation}
where $z_0$ is the $z$-position at abscissa $x_0$ of the trajectory passing through the position $(x,z)$. Finally, the conditional PDF $p_z$ is simply given by $p_z = p/\int p(x,z) dz$. 

\begin{figure}[] 
\includegraphics[width=\textwidth]{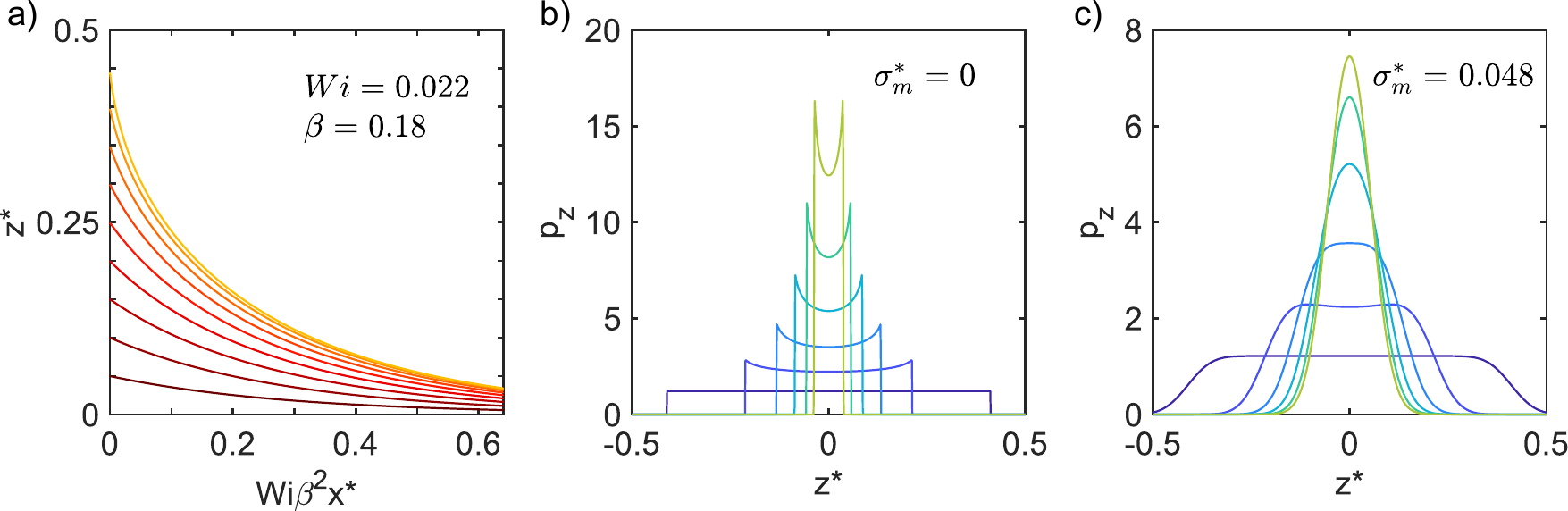}
\caption{a) Theoretical particle trajectories. b) Theoretical PDF evolution along the channel. c) Theoretical prediction, after convolution with a Gaussian function whose standard deviation $\sigma_m^*$ is representative of the uncertainty of the $z$ measurement. PDF are represented for $\mathrm{Wi}\beta^2x^*\in\left[0,0.122,0.244,0.366,0.488,0.61\right]$} \label{FigPDFth}
\end{figure}

Assuming a uniform particle distribution at the inlet, we calculate using this approach the PDF $p_z$ for several positions $x$. An example is displayed in Fig.~\ref{FigPDFth}, together with the corresponding trajectories. The particle trajectories, and thus the PDFs, only depend on the parameter $\text{Wi} \beta^2$, which can be used to rescale the longitudinal axis. The focusing of the particles is quite fast at the beginning and the theoretical PDFs exhibit two symmetric and sharp maxima, corresponding to the extreme trajectories of particles initially located at the walls. The experimental data do not exhibit such maxima, which we interpret as a consequence of the measurement uncertainty. Indeed, when the theoretical PDFs are convolved with a Gaussian function of standard deviation 0.048 (which is the measurement relative uncertainty), these maxima are hardly visible, as shown in Fig.~\ref{FigPDFth}c. The global shapes of the PDFs are very similar to the measured one (Fig.~\ref{FigPDF}).



We use the procedure described in the following, applied to all the experimental conditions. In order to remove any uncertainty coming from the initial particle distribution, we use the first PDF (at $x=1$~cm) as the reference one, and deduce the other PDFs using Eq.~\ref{EqTrajTh} and \ref{EqPz}. It is crucial to take into account the uncertainty of the measurement, which is not negligible as compared to the PDF standard deviation. We assume that measured PDFs result from the convolution of the theoretical PDFs with a Gaussian function of standard deviation 0.048. As the deconvolution operation on experimental data is not easy and requires a strong smoothing, we proceed by convolving the theoretical PDFs. Then, all the experimental PDFs obtained at a given pressure drop and at different $x$ are fitted simultaneously by the convolved theoretical PDFs, with a single fitting parameter K. The agreement is in all cases very good, as could be seen in the examples displayed in Fig.~\ref{FigPDF}. The best fit to the data is obtained with values of $K$ of about 10, ranging from 5 to 16, depending on the experiments. More precisely, the mean value is 10.91, and the standard deviation is 2.76.

\begin{acknowledgments}
This work was supported by the ANR grant $\mathrm{\mu}$LAS (ANR-16-CE18-0028-01). LRP is part of the LabEx Tec21 (ANR-11-LABX-0030) and of the PolyNat Carnot Institute (ANR-11-CARN-007-01). 
	\end{acknowledgments}

\bibliography{BiblioArticles}

\end{document}